\documentclass[
preprint,
showpacs,
amsmath,amssymb,
aps,pre,
]{revtex4-1}
\usepackage{booktabs}
\usepackage{tabularx}
\usepackage{float}
\usepackage{graphicx}
\usepackage{bm}
\usepackage{subfigure}
\usepackage{color}
\usepackage{tikz}
\newcommand*{\circled}[1]{\lower.7ex\hbox{\tikz\draw (0pt, 0pt)%
    circle (.5em) node {\makebox[0em][c]{\small #1}};}}

\usepackage[colorlinks,citecolor = blue, linkcolor=black,hyperindex,CJKbookmarks]{hyperref}

\newcommand{\Ma}{\mathit{Ma}}
\newcommand{\Sc}{\mathit{Sc}}

\begin{document}
\title{Droplet plume emission during plasmonic bubble growth in ternary liquids}

\author{Xiaolai Li$^{1,2}$}
\author{Yibo Chen$^{1}$}

\author{Yuliang Wang$^{2,3}$}\email{wangyuliang@buaa.edu.cn}
\author{Kai Leong Chong$^{1}$}
\author{Roberto Verzicco$^{1,4,5}$}
\author{Harold J. W. Zandvliet$^{6}$}\email{h.j.w.zandvliet@utwente.nl}
\author{Detlef Lohse$^{1,7}$}\email{d.lohse@utwente.nl}

\affiliation{$^1$Physics of Fluids, Max Planck Center Twente for Complex Fluid Dynamics and J.M. Burgers Centre for Fluid Mechanics, MESA$^+$ Institute, University of Twente, P.O. Box 217, 7500AE Enschede, The Netherlands\\
$^2$School of Mechanical Engineering and Automation, Beihang University, 37 Xueyuan Rd, Haidian District, Beijing, China\\
$^3$Beijing Advanced Innovation Center for Biomedical Engineering, Beihang University, 37 Xueyuan Rd, Haidian District, Beijing, China\\
$^4$Dipartimento di Ingegneria Industriale, University of Rome 'Tor Vergata', Roma 00133, Italy\\
$^5$Gran Sasso Science Institute - Viale F. Crispi, 7 67100 L'Aquila, Italy\\
$^6$Physics of Interfaces and Nanomaterials, MESA+ Institute, University of Twente, P.O. Box 217, 7500 AE Enschede, The Netherlands\\
$^7$Max Planck Institute for Dynamics and Self-Organization, 37077 G\"ottingen, Germany\\.}


\begin{abstract}
Plasmonic bubbles are of great relevance in numerous applications, including catalytic reactions, micro/nanomanipulation of molecules or particles dispersed in liquids, and cancer therapeutics. So far, studies have been focused on bubble nucleation in pure liquids.
Here we investigate plasmonic bubble nucleation in ternary liquids consisting of ethanol, water, and trans-anethole oil which can show the so-called ouzo-effect. We find that oil (trans-anethole) droplet plumes are produced around the growing plasmonic bubbles.
The nucleation of the microdroplets and their organization in droplet plumes is due to the symmetry breaking of the ethanol concentration field during the selective evaporation of ethanol from the surrounding ternary liquids into the growing plasmonic bubbles.
Numerical simulations show the existence of a critical Marangoni number $\Ma$ (the ratio between solutal advection rate and the diffusion rate), above which the symmetry breaking of the ethanol concentration field occurs, leading to the emission of the droplet plumes. The numerical results agree with the experimental observation that more plumes are emitted with increasing ethanol-water relative weight ratios and hence $\Ma$.
Our findings on the droplet plume formation reveal the rich phenomena of plasmonic bubble nucleation in multicomponent liquids and help to pave the way to achieve enhanced mixing in multicomponent liquids in chemical, pharmaceutical, and cosmetic industries.
\end{abstract}
\maketitle
\section{Introduction}
Plasmonic microbubbles are produced by immersed metal nanoparticles under laser irradiation due to plasmonic heating. They have received considerable attention, owing to their potential for numerous applications, such as biomedical diagnosis, cancer therapy \cite{liu2014, fan2014, shao2015, emelianov2009, lapotko2009}, micromanipulation of molecules or particles dispersed in solutions \cite{zhao2014, xie2017opt, Tantussi2018} and locally enhanced chemical reactions \cite{adleman2009, baffou2014}. A proper understanding of the dynamics of these plasmonic bubbles is therefore not only of fundamental interest, but also essential from a technological viewpoint as it might provide routes to tailor the formation and growth of the plasmonic bubbles.

During the last decade, numerous studies have been performed to explore the nucleation mechanism and growth dynamics of plasmonic bubbles \cite{lombard2014, Lombard2015, Mahe2018, Lombard2016, wang2018, jollans2019, hou2015explosive}, as well as the effects of dissolved gas \cite{zeng2020, Zaytsev2020, wang2017, li2019, baffou2014jpc} and liquid types \cite{Zaytsev2018} on bubble growth dynamics. To date, these studies have
nearly exclusively focused on plasmonic bubble formation in pure liquids. However, in many applications plasmonic bubbles nucleate in binary, ternary or even multicomponent liquids. This introduces new and rich physics and chemistry. E.g., due to selective evaporation, concentration gradients may occur, resulting in Marangoni forces and other rich physicochemical hydrodynamics \cite{levich1962, lohse2020}, which we want to further explore in this paper.

We have recently started this line of research and investigated plasmonic bubble nucleation in \textit{binary} liquids of ethanol and water, and found that the solutal Marangoni flow, caused by the preferential vaporization of ethanol, results in the detachment of the bubbles from the substrate at low ethanol ratios ($<67.5\%$) \cite{li2020jpcl}. In binary liquids with ethanol ratios exceeding $67.5\%$, on the contrary, bubbles remain attached to the substrate and first steadily grow until they suddenly shrink. The shrinkage is induced by the depinning of the three phase contact line from the laser spot region, which is due to the decreased opto-thermal efficiency.
For plasmonic bubbles in ternary liquids, the phase equilibrium is continuously altered by liquid vaporization and mass and thermal diffusion, resulting in a much richer and more complex behavior than in pure liquids.

To gain insight into the dynamics of plasmonic bubbles in ternary and multicomponent liquids, we study here the bubble formation in a controlled specific category of ternary liquids. Namely, we have selected a ternary liquid which shows the so-called ouzo effect \cite{vital2003, tan2016pnas, lohse2020}, namely a solution consisting of ethanol, water, and trans-anethole. The ouzo effect describes the spontaneous emulsification that occurs when water is added to dilute the solution, thus lowering its oil-solubility. The initially transparent solution then becomes milky because of the nucleation of oil microdroplets due to the reduced solubility of the oil in liquid mixtures with lower ethanol concentration \cite{ganachaud2005, vital2003}. The emulsification can also occur if the amount of ethanol is reduced in different ways, for instance, through preferential evaporation or dissolution \cite{tan2017sf, tan2019jfm, tan2016pnas}.

In this work, we present experimental and numerical studies on plasmonic bubble formation in the ternary liquids of ethanol, water , and trans-anethole. Interestingly, for certain ternary liquid ratios, we observe the emission of droplet plumes from the growing plasmonic bubbles, carrying the nucleated oil microdroplets away from the bubble. We will show that the key parameter governing the pattern and number of these emitted plumes is the Marangoni number, which is the dimensionless number that compares the solutal advection rate with the diffusion rate.  By considering the experiments, numerical simulation and theory, we can provide a detailed understanding of the physicochemical processes that occur during plasmonic bubble growth in these multicomponent liquids.

Note that the formation of trans-anethole droplets out of an  oversaturated ternary solution is an interesting nucleation problem in itself. It could be addressed from the viewpoint of classcial nucleation theory (CNT) \cite{kashchiev2000, ayuba2018, vehkamaki2006}, but this is not the
subject of the  present paper.
If one were to do so, one would  choose  a much simpler geometry, say, by injecting  water
into an ouzo solution of well-defined properties. We do not have this here, due to partial evaporation of the binary liquid
and thermal effects.  So here it would be impossible to determine the values of the material parameters required in classical
nucleation theory. In addition, it is nearly impossible to identify the size of individual droplets or their numbers in the droplet plumes, due to the limited image resolution of the high-speed imaging system,  as well as the cloudy nature of the plumes.
In a nutshell, the cloud of nucleated microdroplets is used as a convenient indicator to signal flow regions of relatively high water
concentration or equivalently  low ethanol concentration. In fact, we will show that the plume formation also works around
plasmonic bubble in
binary liquids, and can optically be visualized for large enough density contrast of the two involved liquids.

The paper is organized as follows:
In section \ref{method}, we describe the experimental setup and procedure. Section \ref{T-exp} reports the experimental results for the plume
formation around the growing plasmonic bubble in a ternary liquid and section \ref{T-B} compares them with those in a binary liquid.
In section \ref{Numerical}, we present our corresponding numerical results, which, in section \ref{Com-Ma}, are quantitatively compared with the
experimental ones. The paper ends with conclusions and an outlook (section \ref{conclusion}).

\section{Experimental setup and procedure}\label{method}
\subsection{Sample Preparation }
\begin{figure*}[hbp]
\begin{center}
 	\includegraphics[width=0.8\textwidth]{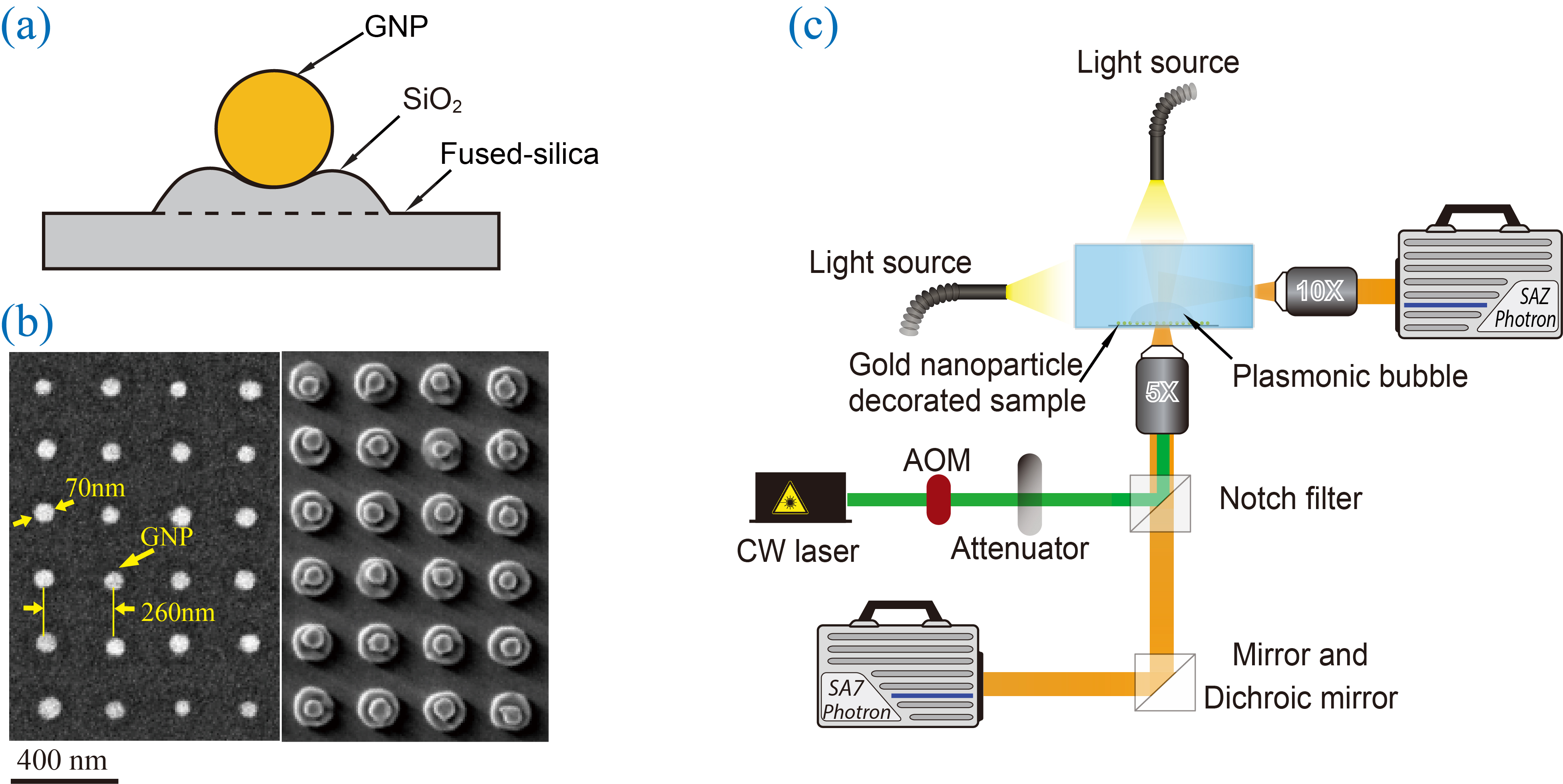}
 	\caption{(a) Schematic of a gold nanoparticle sitting on a SiO$_{2}$ island on a fused-silica substrate, (b) A energy-selective backscatter (left) and scanning electron microscope (right) image of the patterned gold nanoparticle sample surface, (c) Schematic diagram of the optical setup for plasmonic microbubble imaging.
 	}
 	\label{setup}
 \end{center}
 \end{figure*}
A fused silica surface patterned with an array of gold nanoparticles was used to produce plasmonic bubbles. A gold layer of approximately 45 nm was first deposited on an amorphous fused-silica wafer by using an ion-beam sputtering system (home-built T$^\prime$COathy machine, MESA$^+$ NanoLab, Twente University). A bottom anti-reflection coating (BARC) layer ($\sim$186 nm) and a photoresist (PR) layer ($\sim$200 nm) were subsequently coated on the wafer. Periodic nanocolumns with diameters of approximately 110 nm were patterned in the PR layer using displacement Talbot lithography (PhableR 100C, EULITHA). These periodic PR nanocolumns were subsequently  transferred at wafer level to the underlying BARC layer, forming 110 nm BARC nanocolumns by using nitrogen plasma etching (home-built TEtske machine, NanoLab) at 10 mTorr and 25 W for 8 min. Using these BARC nanocolumns as a mask, the Au layer was subsequently etched by ion beam etching (Oxford i300, Oxford Instruments, United Kingdom) with 5 sccm Ar and 50-55 mA at an inclined angle of $5^{\circ}$. The etching for 9 min resulted in periodic Au nanodots supported on cone-shaped fused-silica features. The remaining BARC was stripped using oxygen plasma for 10 min (TePla 300E, PVA TePla AG, Germany). The fabricated array of Au nanodots was heated to $1100 ^{\circ}$C in 90 min and subsequently cooled passively to room temperature. During the annealing process, these Au nanodots re-formed into spherical-shaped Au nanoparticles. Figure \ref{setup}a shows the schematic of a gold nanoparticle sitting on a SiO$_2$ island on a fused-silica. The energy-selective backscatter (ESB) and scanning electron microscope (SEM) images of the patterned gold nanoparticle sample surface are shown in Figure \ref{setup}b (left) and Figure \ref{setup}b (right), respectively.
\subsection{Ternary liquid preparation}
The ternary liquids used in our experiments are composed of ethanol (Sigma-Aldrich; $\geq$ 99.8$\%$), deionized (DI) water (Milli-Q Advantage A10 System, Germany), and trans-anethole oil (Sigma-Aldrich; trans-anethole, $\geq$99.8). The ethanol and water were air saturated by exposure to air for 8 hours. Subsequently, they were mixed with trans-anethole oil to prepare seven different ternary liquids, as shown in Table \ref{table1}. After mixing, the ternary liquids were immediately transferred into a sealed cuvette with a size of 10mm $\times$ 10mm $\times$ 50mm for experiments. During experiments, the cuvette was fully filled with the ternary liquid.
\subsection{Setup description}
The experimental setup for plasmonic microbubble imaging is shown in Figure \ref{setup}c. The gold nanoparticle decorated sample was placed in a quartz glass cuvette and filled with ternary liquid. A continuous-wave laser (Cobolt Samba) of 532 nm wavelength with a maximum power of 300 mW was used for sample irradiation. An acousto-optic modulator (Opto-Electronic, AOTFncVIS) was used as a shutter to control the laser irradiation on the sample surface. A pulse/delay generator (BNC model 565) was used to generate two different laser pulses of 400 $\mu s$ and 4 s in order to study the short-term and the long-term dynamics of microbubbles, respectively. The laser power was controlled by using a half-wave plate and a polarizer and measured by a photodiode power sensor (S130C, ThorLabs). Two high-speed cameras were installed in the setup. One (Photron SA7) was equipped with a 5x long working distance objective (LMPLFLN, Olympus) and the other (Photron SAZ) is equipped with various long working distance objectives: 10$\times$ (LMPLFLN, Olympus) and operated at frame rate of 5 kfps. The first camera was used for top-view imaging and the second one was for side-view imaging. Two light sources, Olympus ILP-1 and Schott ACE I were applied to provide illumination for the two high-speed cameras.
The optical images are processed with a home designed image segmentation algorithm for the optimized extraction of the bubble radius in MATLAB.

\subsection{Experimental procedure}
\begin{figure}[h]
\centering
  \includegraphics[width=0.5\textwidth]{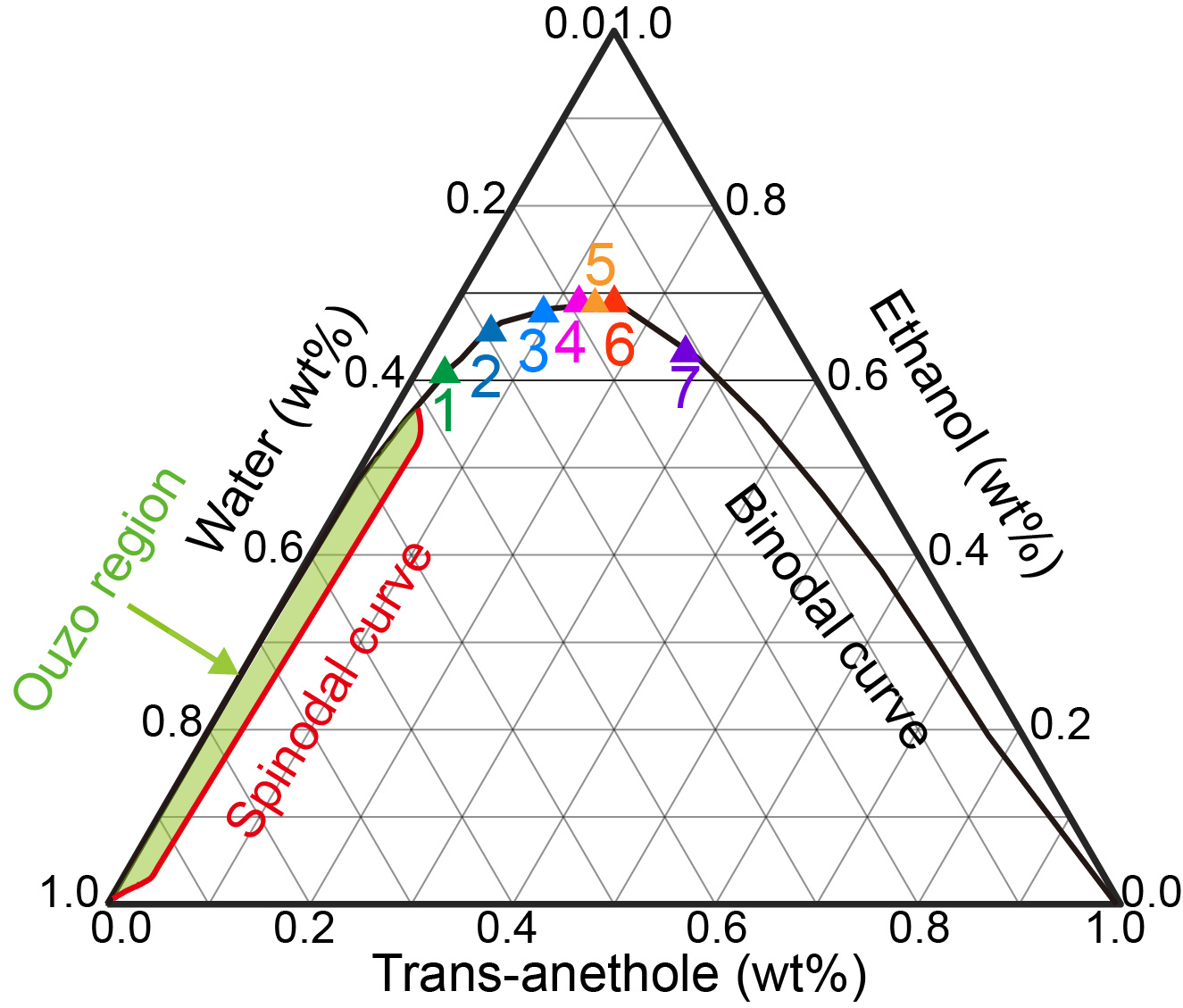}
  \caption{Ternary diagram of ethanol-water-anethole mixtures. The black solid line is the measured phase separation line\cite{tan2017sf}. The filled triangles (points 1-7) along the phase separation curve represent the initial compositions of the ternary liquids that we have used in this study.}
  \label{diagram}
\end{figure}

Plasmonic bubble generation experiments were performed in equilibrated ethanol-water-anethole mixtures. As depicted in the ternary diagram (Figure \ref{diagram}), ternary liquids with seven different weight ratios, labelled from 1 to 7, were selected on the binodal curve (phase separation curve) for our experiments. The phase separation curve is based on the study by Tan \textit{et al} \cite{tan2016pnas}. In the ternary mixtures, the weight ratios of ethanol, water, and anethole oil are $y_e$, $y_w$, and $y_a$, respectively. Table \ref{table1} summarizes the detailed weight ratios of each component in the ternary liquid mixtures. Note that the absolute ethanol concentration $y_e$ non-monotonically changes along the binodal curve. The seven ternary liquids were actually labelled with increasing ethanol-water relative weight ratio of $r_{e/(e+w)}=y_e/(y_e+y_w)$, which hereafter is used for analysis. At low ethanol concentrations, the plasmonic bubbles depin and detach due to the strong upward Marangoni force, which has been systematically investigated in our previous study \cite{li2020jpcl}. In contrast, bubbles can steadily attach to the substrate at high ethanol fractions.

\begin{table}[h]
 \caption{Weight ratios of $y_e$, $y_w$, $y_a$ for ethanol, water, and anethole oil in different ternary mixtures used for plasmonic bubble formation. The liquids were labelled with increasing ethanol-water relative weight ratio $r_{e/(e+w)}$.}
 \centering
  \begin{tabular}{ccccc}
    \hline
    Label & $y_e(wt\%)$ & $y_w(wt\%)$ & $y_a(wt\%)$ & $r_{e/(e+w)}=y_e/(y_e+y_w$) \\
    \hline
    1 & 62.58  & 33.71 & 3.71 & 65.0\% \\
    2 & 66.31  & 28.42 & 5.27 & 70.0\%\\
    3 & 68.09  & 22.70 & 9.21 &  75.0\%\\
    4 & 68.56  & 19.90 & 11.54 &  77.5\%\\
    5 & 68.67  & 18.25 & 13.07 & 79.0\%\\
    6 & 68.66  & 17.17 & 14.17 & 80.0\%\\
    7 & 64.15  & 11.32 & 24.52 & 85.0\%\\
    \hline
  \end{tabular}
  \label{table1}
\end{table}

\section{Plume formation around the growing plasmonic bubble in a  ternary liquid}\label{T-exp}
\begin{figure}[htbp]
\centering
  \includegraphics[width=0.8\textwidth]{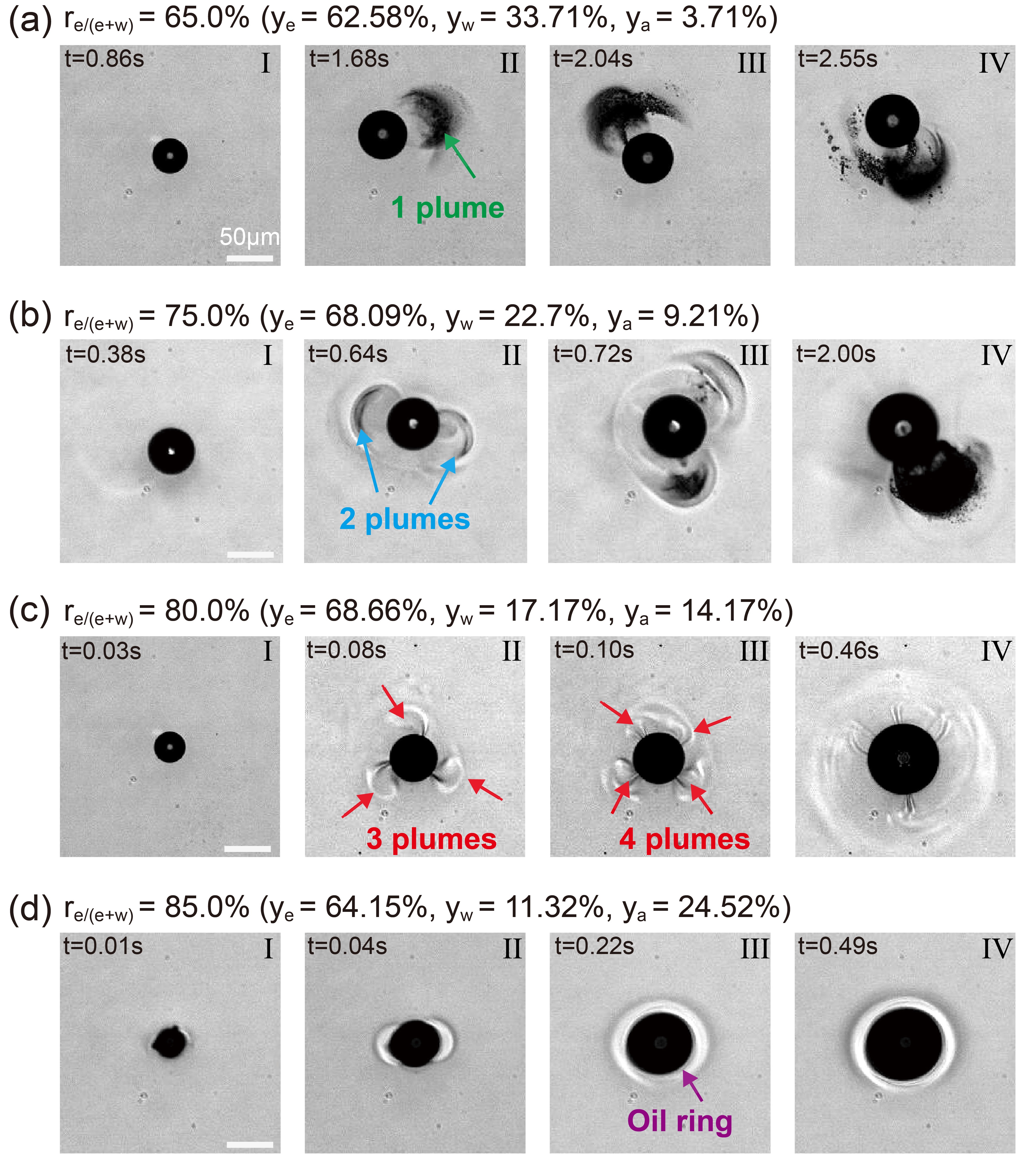}
  \caption{Bottom-view images of droplet plumes during plasmonic microbubble growth in ternary liquids. (a), (b), (c) and (d) show the evolution of droplet plumes with ethanol-water relative weight ratio $r_{e/(e+w)}=65.0\%$, 75.0\%, 80.0\%, and 85.0\%, respectively. With increasing $r_{e/(e+w)}$ from 65.0\% to 80.0\%, the initially emitted plume number increases from 1 to 3, as seen in Figure \ref{exp}a II, Figure \ref{exp}b II, and Figure \ref{exp}c II. For $r_{e/(e+w)} = 75.0\%$, the two plumes finally merge into a single plume (Figure \ref{exp}b II-IV). For $r_{e/(e+w)}=80.0\%$, the initial three plumes (Figure \ref{exp}c II) evolve into four plumes (Figure \ref{exp}c III) and finally merge (Figure \ref{exp}c IV). Figure \ref{exp}d shows the formation of an oil ring in the ternary liquid with $r_{e/(e+w)}=85.0\%$, corresponding to a high oil weight fraction $y_a = 24.52\%$.}
  \label{exp}
\end{figure}

We first report our finding for the  second regime, \textit{i.e.}, the regime with  high ethanol fractions.
Figure \ref{exp}a (I) shows a bottom-view image of a growing plasmonic bubble in a ternary liquid with ethanol-water relative weight ratio $r_{e/(e+w)}=65.0\%$, corresponding to ethanol $y_e = 62.58\%$, water $y_w = 33.71\%$, and trans-anethole $y_a = 3.71\%$ in weight fractions. Interestingly, a cloudy black region consisting of oil microdroplets is observed starting at 1.68s (Figure \ref{exp}a II). The emulsification here results from the selective evaporation of ethanol from the ternary liquid \cite{vital2003}. Since ethanol has the lowest boiling temperature of the three components in the ternary liquid, ethanol vaporizes first, leading to an oversaturation of the left-behind liquid with the anethole oil and thus to the nucleation of anethole oil microdroplets. Remarkably, the cloudy region keeps rotating around the bubble. The bubble also displays some lateral movement (Figure \ref{exp}a III and IV). Hereafter we refer to the localized cloudy microdroplet regions as droplet plumes.

The formation of the droplet plumes is also observed in ternary liquids with the ethanol-water relative weight ratio of $r_{e/(e+w)}= 70.0-80.0\%$, as shown in Figure \ref{exp}b-d. For $r_{e/(e+w)}=70.0\%$, we observe the emission of two droplet plumes (Figure \ref{exp}b II). The two plumes are formed at opposite sides of the bubble. Subsequently, they merge into a single droplet plume (Figure \ref{exp}b III and IV). For $r_{e/(e+w)}=80.0\%$, there are initially three plumes (Figure \ref{exp}c II), which evolve later into four plumes (Figure \ref{exp}c III). Similar to the case of $r_{e/(e+w)}=70.0\%$, the plumes are first distributed symmetrically around the bubble until eventually the localized plumes merge (Figure \ref{exp}c IV).
However, in the ternary liquids with $r_{e/(e+w)}=85.0\%$ (Figure \ref{exp}d), there is no pronounced droplet plume observed, but rather an oil ring is formed around the bubble (Figure \ref{exp}d II-IV). This can be rationalized by the high oil concentration ($y_a = 24.52\%$) dissolved in the original ternary liquids. Once the ethanol evaporates, the nucleated microdroplets accumulate around the bubble, coalesce and form the oil ring.
Note that the bubble formation experiments were performed in a cuvette (10mm $\times$ 10mm $\times$ 50mm) fully filled with the ternary liquids. Therefore, the thickness of liquid inside the cuvette is about 10 mm, which is nearly 100 times larger than the bubble size ($\sim$100 $\mu$m). The influence of the wall and interface on convection is negligible.

\begin{figure}[htbp]
\centering
  \includegraphics[width=0.9\textwidth]{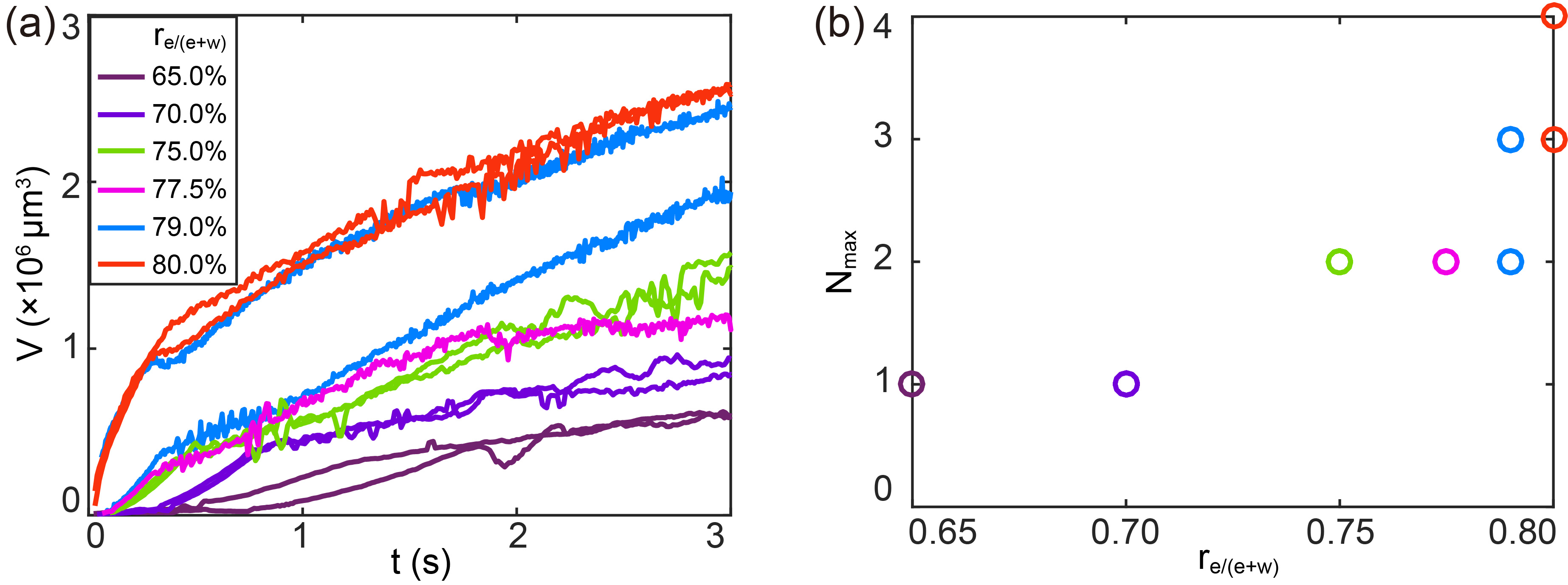}
  \caption{(a) Bubble volume $V$ versus time and (b) maximum plume number $N_{max}$ versus ethanol-water relative weight ratio $r_{e/(e+w)}$ in the ternary liquids. }
  \label{volume}
\end{figure}

To get more insight into the bubble dynamics and the droplet plume formation, the bubble volume and maximum plume number for different liquid ratios were extracted from our high-speed images. In Figures \ref{volume}a and b we plotted the volume of the plasmonic bubble versus time and the maximum number of droplet plumes versus the ethanol ratio $r_{e/(e+w)}$.
Not surprisingly, the bubble growth in the ternary liquids here is very comparable to the bubble growth in ethanol-water binary liquids \cite{li2020jpcl}.
As shown in Figure \ref{volume}a, for fixed laser irradiation time, the bubble volume increases with increasing ethanol-water relative weight ratio $r_{e/(e+w)}$ from $65.0\%$ to $80.0\%$. The anethole oil ratio in our ternary system is very low and anethole oil has a much higher boiling temperature than ethanol and water. As a result, anethole oil has limited influence on the vaporization process and the bubble growth in the ternary liquids. A larger relative ethanol fraction in ethanol-water binary liquids results in a large bubble growth rate \cite{li2020jpcl}. The maximum plume number, $N_{max}$, was found to increase with increasing $r_{e/(e+w)}$, see Figure \ref{volume}b.

We note that the two curves at $r_{e/(e+w)} = 79.0\%$ deviate from each other in Figure \ref{volume}a. We believe that this is due to the fact that $r_{e/(e+w)} = 79.0\%$ is close to the critical ethanol fraction (77.5$\%$-80.0$\%$) at which the transition from depinning (at $r_{e/(e+w)} < 77.5\%$) to pinning ($r_{e/(e+w)} \geq 80.0\%$, see Figure \ref{exp}c) of the plasmonic bubble to the laser spot occurs. For a high ethanol ratio $r_{e/(e+w)} \geq 80.0\%$, the bubble remains pinned to the laser spot region (Figure \ref{exp}c), the position of which locates at the center of each image in Figure \ref{exp}. As a result, one can barely see the lateral movement of the bubbles for $r_{e/(e+w)} \geq 80.0\%$. However, for a lower ethanol ratio $r_{e/(e+w)} < 77.5\%$, bubbles are not pinned at the laser spot (see Figures \ref{exp}a and b). They could move away from the laser spot region. The laser spot has a Gaussian power density distribution, so the further the bubbles are located away from the laser spot center, the weaker the heating efficiency as well as the vaporization rate. As a result, apart from the dependence on the ethanol ratio $r_{e/(e+w)}$, the growth rate of those bubbles that are depinned from the laser region is even smaller.

The mechanism of the droplet nucleation in experiments can be explained as follows. Under continuous laser heating the volatile ethanol in ternary liquids vaporizes into the bubble. Consequently, the liquid in the region near the bubble interface becomes ethanol depleted, and an ethanol concentration gradient emerges. The reduction of ethanol in the initially equilibrated ternary liquid leads to the oversaturation of the left-behind liquid with the anethole oil, and eventually to the nucleation of oil microdroplets (the so-called ouzo effect). However, the reason why the symmetry of oil droplet distribution around the bubble breaks and how the droplet plumes form remains to be explained, and we will do so below.

\section{Comparison of plume formation around  growing plasmonic bubbles in binary and in ternary liquids}\label{T-B}
\begin{figure*}[h]
\begin{center}
 	\includegraphics[width=0.8\textwidth]{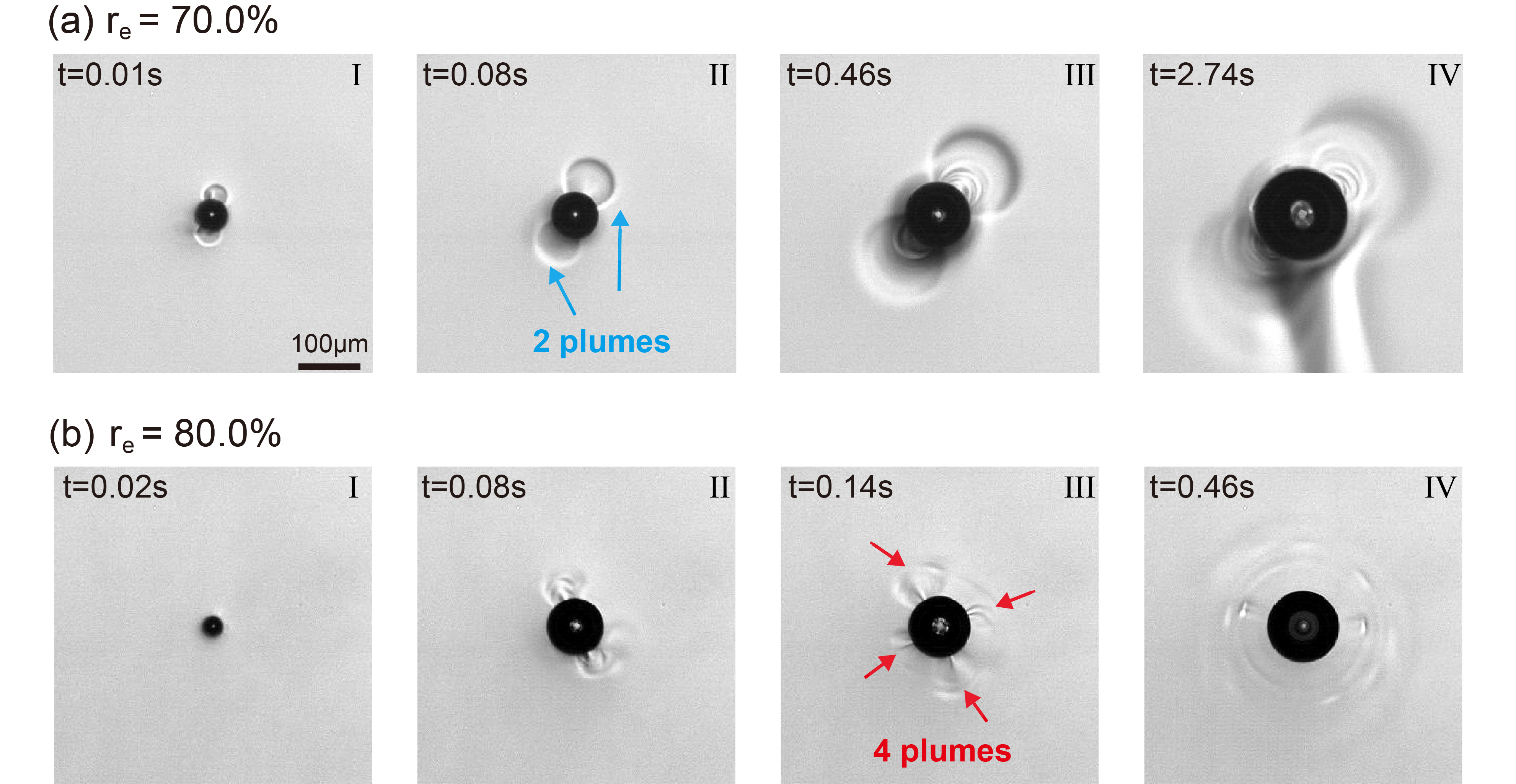}
 	\caption{Plasmonic bubbles in ethanol and anethol oil mixtures with ethanol ratio $r_e=70\%$ (a) and $r_e=80\%$ (b). Plume formation is observed in both cases and the maximum plume number increases with increasing ethanol ratio.
 	}
 	\label{binary}
 \end{center}
 \end{figure*}

To demonstrate that the oil droplets do not change the essence of the overall flow dynamics and to provide a comparison of the cases with and without oil microdroplet
nucleation, we   performed another experiment in a binary liquid of ethanol and trans-anethol oil, which are fully miscible.
We observed a similar plume phenomenon, as shown in Figures \ref{binary}a and b. Ethanol and anethol oil can be mixed in any ratio. During bubble growth under laser heating, it is ethanol which preferably vaporizes, due to the ethanol low boiling point. As a result, a concentration gradient of anethol oil in the vicinity of the laser irradiated area will form and can be visualized
 because of the different refractive index of ethanol and oil.

Figures \ref{binary}a and b show the plasmonic bubble growth in ethanol anethol mixtures with ethanol ratio $r_e=70\%$ and 80\%, respectively. It can be seen that during bubble growth, plumes are generated in the ethanol anethol oil mixtures, and the maximum plume number increases with increasing ethanol ratio. This
 verifies that the plumes can be formed regardless of the droplet nucleation.

For the droplet plumes in ternary liquids of ethanol, water and trans-anethol, the hydrodynamics of the plumes is coupled with the droplet nucleation. We note that there are certain quantitative differences between the experiments in ternary liquids (Figure \ref{exp}) and in binary liquids (Figure \ref{binary}).  This might be attributed to the influence of phase nucleation in the ternary liquids, in which an energy barrier must be surmounted to bring the system in visible stable phase transition. Moreover, the phase nucleation process is controlled by thermal fluctuations and affected by the defects of wall or impurities in liquids, leading to the stochastic properties of droplet nucleation with ternary liquids.  However, a quantitative analysis of these differences
is beyond the scope of this study.

\section{Numerical simulations of the plasmonic bubble in an ethanol-water mixture}\label{Numerical}
\subsection{Underlying equations and control parameters}
\begin{figure}[htbp]
\centering
  \includegraphics[width=0.8\textwidth]{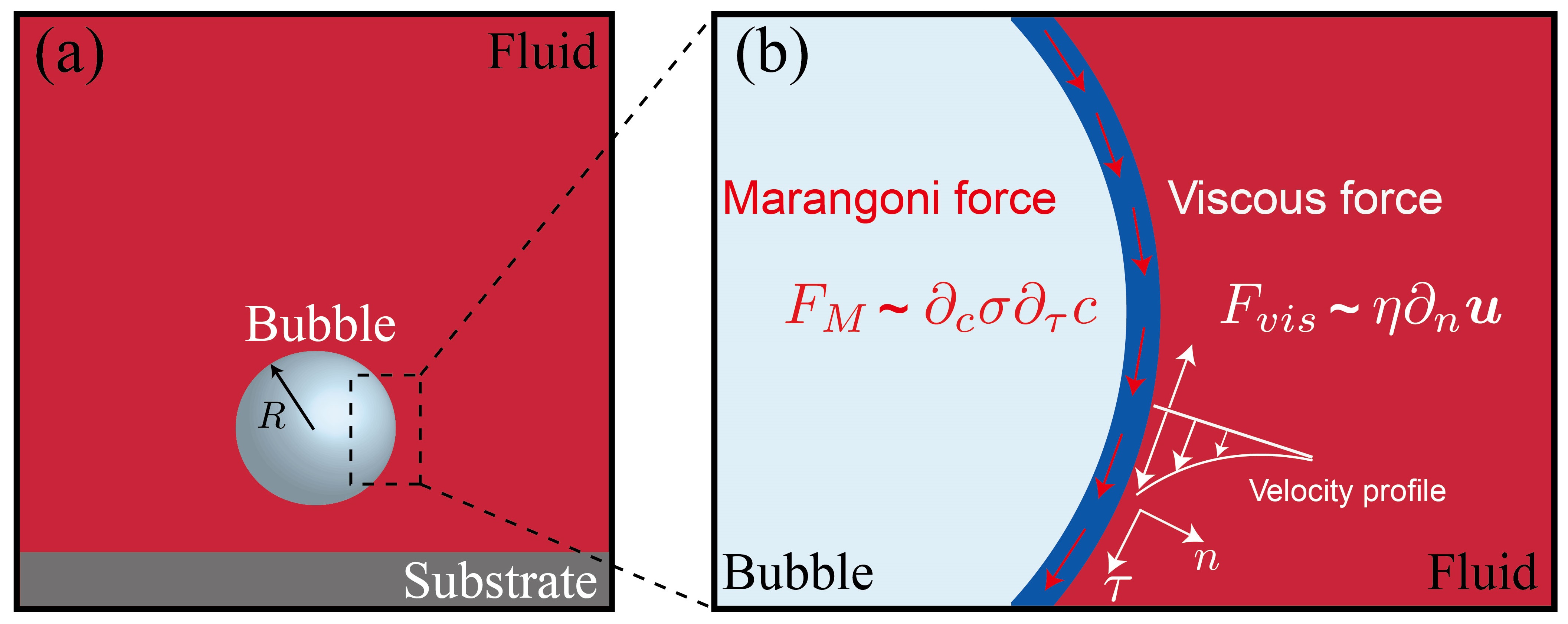}
  \caption{(a) Schematics of the numerical model. The domain is filled with the ethanol-water mixture and a bubble of radius $R$ is placed near the bottom wall. (b) A zoom-in display for the area selected by the dashed box at the bubble surface in (a). Concentration differences along the bubble surface lead to a Marangoni force, which is balanced by the viscous force induced by the Marangoni flow.}
  \label{sketch}
\end{figure}

To confirm the outlined physical picture, we numerically investigate the vaporization and plume formation during plasmonic bubble growth. Since the ratio of anethole oil in the ternary liquid is very low and the boiling point of anethole oil (234 $^\circ$C) is much higher than that of water and ethanol, it hardly vaporizes during the experiments. Therefore, to reduce the complexity of the numerical model, we only consider water and ethanol. The physical variables to describe the system are thus the concentration of ethanol $c(\boldsymbol{x},t)$ and the velocity of the fluid $\boldsymbol{u}(\boldsymbol{x},t)$. A bubble of radius $R$ immersed in the ethanol-water mixture is fixed near the bottom wall (Figure \ref{sketch}a).
Under laser irradiation, the bubble takes up ethanol vapor with a constant rate $\alpha$ from the fluid at the bubble surface:
\begin{equation}
D \partial_n c=\alpha.
\label{equ1}
\end{equation}
where $D$ is the diffusion coefficient of ethanol, and $\partial_n c$ is the concentration gradient in normal direction.
The ethanol concentration difference along the bubble surface will lead to a surface tension difference, and induce a Marangoni flow. The solutal Marangoni flow is accounted for by balancing the surface tension difference with the viscous force at the interface (Figure \ref{sketch}b) \cite{sternling1959}:
\begin{equation}
\partial_{c} \sigma \partial_{\tau} c=\eta \partial_{n} \boldsymbol{u},
\label{equ2}
\end{equation}
where $\sigma$ is the surface tension, $\partial_{c} \sigma$ represents the derivative of the surface tension with respect to the concentration, $\partial_{\tau} c$ is the tangential concentration gradient, $\eta$ is the dynamic viscosity, and $\partial_{n} \boldsymbol{u}$ is the normal gradient of the velocity.

We non-dimensionalize the velocities, lengths, and concentrations with $\partial_{c} \sigma\alpha R/\eta D$, $R$, and $\alpha R/D$, respectively. The time evolution of the concentration field $c$ and the velocity field $\boldsymbol{u}$ are obtained by the Navier-Stokes equation and the advection-diffusion equation:
\begin{equation}
\frac{\partial \boldsymbol{u}}{\partial t}+(\boldsymbol{u} \cdot \nabla) \boldsymbol{u}=-\nabla p+\frac{S c}{M a} \nabla^{2} \boldsymbol{u}, \nabla \cdot \boldsymbol{u}=0
\label{equ3}
\end{equation}
\begin{equation}
\frac{\partial c}{\partial t}+\boldsymbol{u} \cdot \nabla c=\frac{1}{M a} \nabla^{2} c,
\label{equ4}
\end{equation}
where $\Sc$ and $\Ma$ are Schmidt number and Marangoni number, respectively. The Schmidt number is the ratio between the kinematic viscosity $\nu$ to the diffusion coefficient $D$,
\begin{equation}
\Sc=\frac{\nu}{D},
\label{equ5}
\end{equation}
The Marangoni number is the ratio between Marangoni flow induced advection and diffusion, reflecting the strength of the Marangoni force:
\begin{equation}
\Ma=\frac{\alpha R^{2} \partial_{c} \sigma}{\eta D^{2}}.
\label{equ6}
\end{equation}
In the numerical simulation, we solve the non-dimensional form of the governing Eqs. (\ref{equ3}) and (\ref{equ4}) and boundary conditions (Eqs. (\ref{equ1}) and (\ref{equ2})). The kinematic viscosity of ethanol-water mixture is $\nu = 2.7\times10^{-6} m^2/s$ and the diffusion coefficient of ethanol in water is $D = 1.23\times$ 10$^{-9}$ m$^2$/s \cite{tan2016pnas,hills2011}. Accordingly, the Schmidt number $\Sc$ is taken as 2220 in the simulation. $\Ma$ was varied from $5\times10^3$ to $5\times10^5$. For details regarding the numerical simulations we refer to the Appendix \ref{num-details}.
\setlength{\parskip}{0.5\baselineskip}
\subsection{Numerical results}
\begin{figure}[htbp]
\centering
  \includegraphics[width=0.65\textwidth]{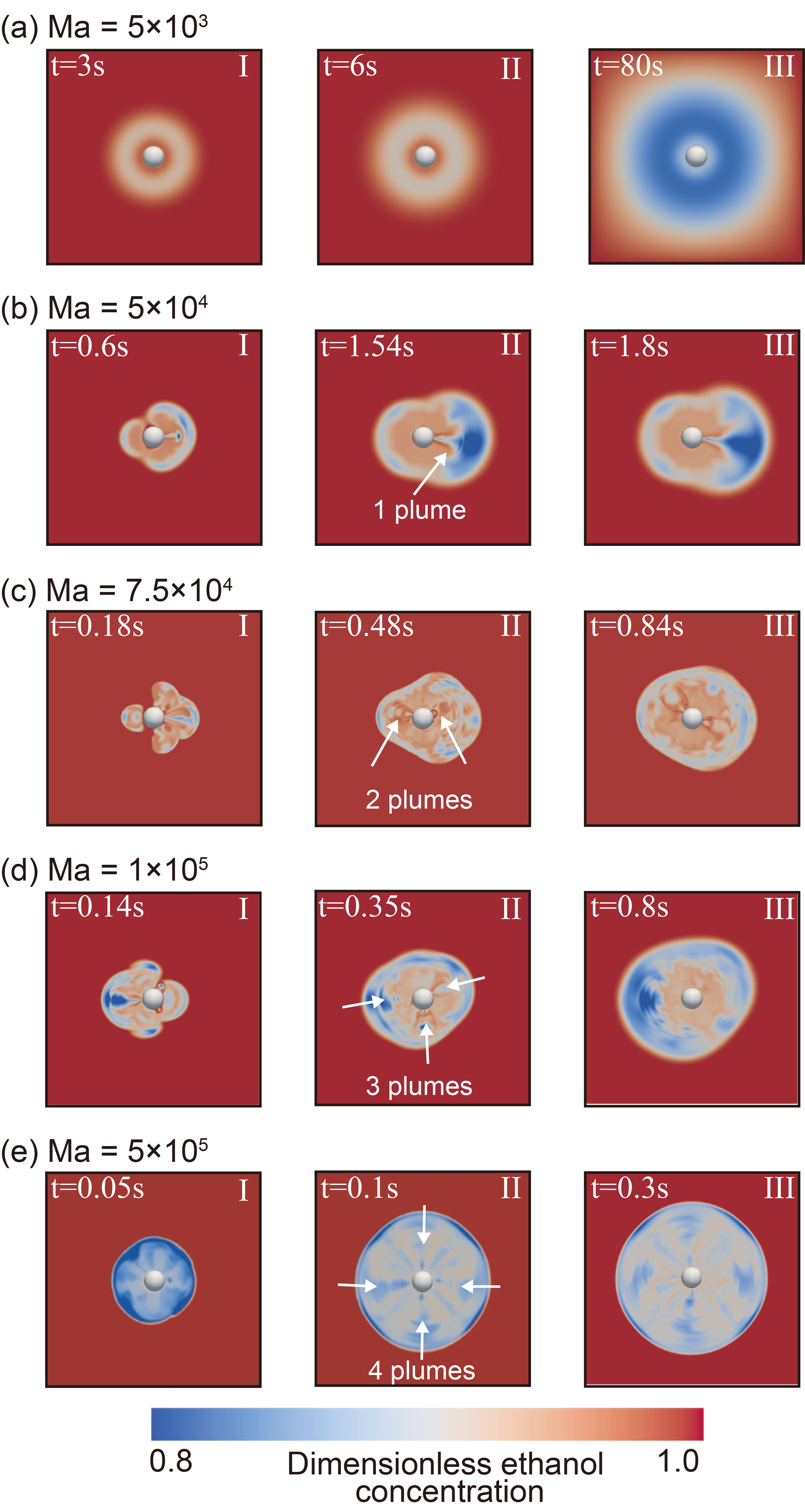}
  \caption{Snapshots (top-view) of the ethanol concentration field near a plasmonic bubble from the simulation in the ethanol water mixtures during the vaporization process. The symmetry of the concentration field breaks for Marangoni numbers $\Ma \geq 5\times 10^4$. With further increasing $\Ma$, the number of plumes increases from 1 to 4. In (c)-(e), neighbouring plumes can merge.}
  \label{sim}
\end{figure}

The simulation results for the ethanol water system with different values of $\Ma$ are shown in Figure \ref{sim}. It is found that the ethanol concentration field remains perfectly symmetric for $\Ma \leq 5\times10^3$ (Figure \ref{sim}a I-III). For $\Ma = 5\times10^4$, the symmetry of the ethanol concentration breaks and a single plume of ethanol poor liquid is emitted (Figure \ref{sim}b). For $\Ma \geq 7.5\times 10^4$, multiple plumes are emitted (Figure \ref{sim}c II, d II and e II). The maximum number of the emitted plumes increases with increasing $\Ma$. Moreover, multiple plumes subsequently merge (Figure \ref{sim}c III, d III and e III), which is consistent with our experimental observations.

The numerical results clearly show that the formation of plumes is directly related to the Marangoni number $\Ma$.
At low Marangoni number $\Ma$, the diffusion process dominates and thus any sharp concentration gradient can be smoothed out by the diffusive process. Therefore, the ethanol concentration field remains symmetric. For $\Ma$ above a certain value, the symmetry of ethanol concentration spontaneously breaks and concentration plumes are formed.

A similar phenomenon of symmetry breaking is observed in the case of catalytic particles with diffusio-phoretic effects. Michelin et al. \cite{Michelin2013} demonstrates that when the P\'eclet number (characterizes the ratio between advection led by diffusio-phoresis and mass diffusion) is above a certain value, the particle will break the symmetry spontaneously.  Chen et al. \cite{yibo2020} further explored catalytic particles at high P\'eclet number. They found that multiple plumes are generated at the surface of the particle and the number of plumes increase with increasing P\'eclet number. The system studied here share similarities with the diffusio-phoretic problem in their symmetry-breaking mechanism and the plume emission at the surface. However, the two systems are different, namely in refs. \cite{Michelin2013, yibo2020} it is the diffusio-phoretic flow at the surface which drives the particle, whereas here it is the Marangoni flow due to the gradient in surface tension.

\section{Quantitative comparison between numerical and experimental results}\label{Com-Ma}
To more quantitatively compare our numerical results with the experimental data, we estimated the Marangoni number $\Ma$ in our experiments with the definition of Eq. (\ref{equ6}). In the experiments, the bubble radii kept changing with time. In order to compare with the numerical results, we take the radius of the bubble at the moment when pronounced plume formation in the experiments is observed to calculate $\Ma$.
As mentioned earlier, we ignored the presence of trans-anethole oil as its relative concentration is very low.
The term $\partial_c\sigma$ is estimated to be 1.7 $\times$ 10$^{-5}$ N$\cdot$m$^2$/kg from the surface tension curves (Appendix \ref{e-w-mix}). To estimate the mass transfer rate $\alpha$, we assume that the bubble growth is dominated by ethanol evaporation. $\alpha$ is defined as $\alpha = K_e(t)/S_b$, where $K_e(t)$ is the consumption rate of ethanol and $S_b$ is the area of bubble surface where the mass transfer takes place.
The consumption rate $K_e(t)$ = d($M_en_e$)/d$t$, where $M_e$ = 46 Kg/mol is the mole weight of ethanol, $n_e = P_0V/R_0T$ is the amount of ethanol molecules in mole, the temperature $T\approx$ 353 K (boiling temperature), $P_0$ the ambient pressure and $R_0$ the ideal gas constant. The consumption rate $K_e(t)$ can be obtained by extracting the bubble volume growth rate, \textit{i.e.} d$V$/d$t$, from the curves shown in Figure \ref{volume}a.

\begin{figure}[hb]
\centering
  \includegraphics[width=1\textwidth]{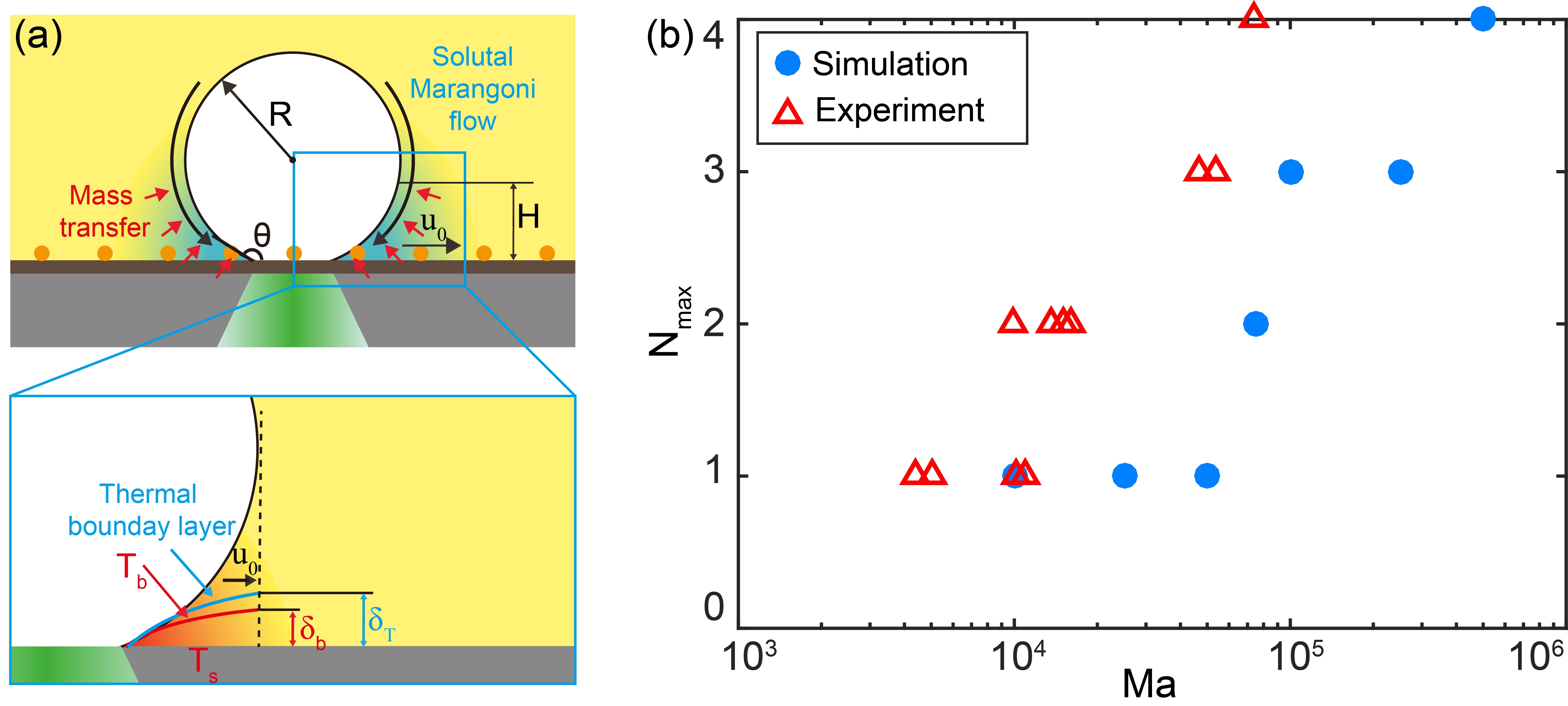}
  \caption{(a) Schematic illustration of the solutal Marangoni flow along the bubble interface caused by selective evaporation of ethanol. Ethanol vaporization occurs in a local area near the bubble bottom. $R$ and $H$ are the bubble radius and the height of the bubble where mass transfer into the bubble can occur, respectively. $u_0$ denotes the flow velocity along the substrate. The zoom-in image shows the thickness $\delta_T$ of the thermal boundary layer. The liquid in the region above the blue curve is at room temperature. The region of thickness $\delta_b$ (below the red curve) has liquid with a temperature above the boiling temperature $T_b$. It can be obtained by assuming a linear temperature decrease from the substrate to the blue curve. (b) Maximum plume number $N_{max}$ as function of the Marangoni number $\Ma$. Theory \textit{vs} experiment.}
\label{comp}
\end{figure}

As depicted in Figure \ref{comp}a, under laser irradiation, only the liquid in the vicinity of the three-phase contact line is heated up to the boiling temperature, resulting in the vaporization of ethanol. Therefore, here we take $S_b = 2\pi R H$ as the bubble surface area for mass transfer. In order to obtain the height $H$, we determine the width of the thermal boundary layer. As shown in Figure \ref{comp}a, there is a downward solutal Marangoni flow along the bubble interface and therefore an outward flow along the substrate due to the selective vaporization of ethanol. This flow can be observed from side view images and the flow velocity $u_0 \approx0.1$ m/s is obtained by particle image velocity measurement. The thermal boundary layer (the blue curve of the zoom in image in Figure \ref{comp}a) is given by $\delta_T  = \delta_vPr^{-1/3}$ \cite{schlichting1961}. In this expression, the velocity boundary layer is given by $\delta_v \approx 5R/Re^{1/2}$, the Reynolds number $Re = u_0x/\nu$, the kinematic viscosity $\nu \approx 2.7\times10^{-6}$ m$^{2}$/s and $x$ is the lateral position. The Prandtl number $Pr = c_p\mu/\lambda$, in which the thermal conductivity $\lambda$ =  0.6 W/mK. The heat capacity of the liquid $c_p = 4179.6$ J$\cdot$kg$^{-1}\cdot$K$^{-1}$.

Room temperature (20 $^{\circ}$C) is taken as $T_0$ for the region outside the thermal boundary layer $\delta_T$  (the blue curve in Figure \ref{comp}a). The substrate temperature $T_s$ is estimated by our numerical simulation by assuming that the substrate is completely covered by vapor. The heat conductivity of the fused silica substrate was taken as 1.4 W/m$\cdot$K (0-20 $^{\circ}$C). The details of the numerical simulation method can be found in our recent publication\cite{wang2018}. Using this method, the spatial-temporal evolution of the substrate temperature can be obtained. The temperature field after 2 s of laser irradiation is shown in Figure \ref{figc5s5} of (Appendix \ref{temp}). The temperature of the substrate at the three phase contact line $L$ is used to determine the thickness of the thermal boundary layer. From the simulation results we obtain a substrate temperature $T_s \approx$ 96 $^{\circ}$C.
By assuming that temperature linearly decreases from the substrate to the blue curve (Figure \ref{comp}a), the liquid region ($\delta _b$) having a temperature above the saturation temperature of ethanol water mixture (about 80 $^{\circ}$C) can be obtained. Our calculation reveals that $\delta _b$ is about 10 $\mu$m at $x = R = 50$ $\mu$m.

The number of plumes as a function of $\Ma$ extracted from the experimental data and the numerical simulations is shown in Figure \ref{comp}b. The numerical and experimental results have the same trend and agree qualitatively well, albeit for a fixed number of plumes the Marangoni number in the simulations is slightly larger than that in the experiments.
We believe that this is due to underestimating $\partial_{c} \sigma$ and the mass transfer rate $\alpha$ in the experimental estimation of $\Ma$. Our arguments for this are twofold:
First, the surface tension gradient $\partial_c\sigma$ was calculated for the initial ethanol concentration. However, the actual ethanol concentration in the boundary layer is much lower than the initial value because of the selective vaporization of ethanol. Since the surface tension of the ethanol-water mixtures changes more rapidly at lower ethanol concentrations than at larger one, the actual value of $\partial_c\sigma$ is larger than that of the initial value.
Second, the ethanol consumption rate $\alpha$ was obtained assuming that the substrate is completely covered by vapor. Under this assumption, the substrate temperature, as well as the thermal boundary layer is determined. However, as depicted in the sketch of Figure \ref{comp}a, part of the substrate is in contact with the liquid outside the pinned bubble footprint area. As the liquid has a higher heat conductivity than vapor, the actual temperature of the substrate must be lower than what we have used in our numerical simulation. The overestimation of the substrate temperature leads to a larger boundary layer and therefore a smaller $\alpha$. These two effects both contribute to the underestimation of the Marangoni numbers from experiments. Nonetheless, it is clear that the numerical results qualitatively agree well with the experimental ones.

\section{Conclusions}\label{conclusion}
To summarize, we have experimentally presented the emulsification of oil droplets induced by plasmonic bubble nucleation in ternary liquids of ethanol-water-anethole. It has been shown that anethole oil microdroplets nucleate around the bubble due to the selective vaporization of ethanol. The oil microdroplets then form droplet plumes. When the ethanol-water relative weight ratio $r_{e/(e+w)}$ increases from 65.0\% to 80.0\% in the ternary liquids, plasmonic bubbles grow faster and more droplet plumes are emitted.
A quantitative understanding on the plume formation is achieved by a numerical study, which shows that the ethanol concentration field can only remain symmetric for a Marangoni number $\Ma<5\times10^4$ (diffusion dominated). Above a critical value of $\Ma=5\times10^4$, a symmetry breaking of the ethanol concentration field takes place, leading to the formation of droplet plumes.
The revealed $\Ma$ dependence of the maximum plume numbers in the numerical results agrees with that in experiments.

This work has given insight into the rich phenomena of plasmonic bubble induced oil microdroplet nucleation (ouzo effect) in ternary liquids and the self-organization of these microdroplets. The quantitative understanding on plume formation offers valuable information for tuning and optimizing the flow field of multiple component liquids in various physicochemical processes. E.g., systems with a high Marangoni number can lead to breaking of the symmetry of the concentration field, and the emerging convection flow may have the advantage of enhanced mixing of ternary liquids in chemical, pharmaceutical, and cosmetic industries.

{\it\textbf{Acknowledgements:}}
The authors greatly appreciate the valuable discussions with Xuehua Zhang. We also thank the Dutch Organization for Research (NWO), The Netherlands Center for Multiscale Catalytic Energy Conversion (MCEC), ERC (via Advanced Grant DDD, Project 740479), and the Chinese Scholarship Council (CSC) for the financial support. Y.W. acknowledges the financial support from NSFC (Grants 51775028 and 52075029) and Beijing Youth Talent Support Program.

\begin{appendix}

\section{Details on the numerical model} \label{num-details}
\begin{figure}[htbp]
\centering
  \includegraphics[width=0.6\textwidth]{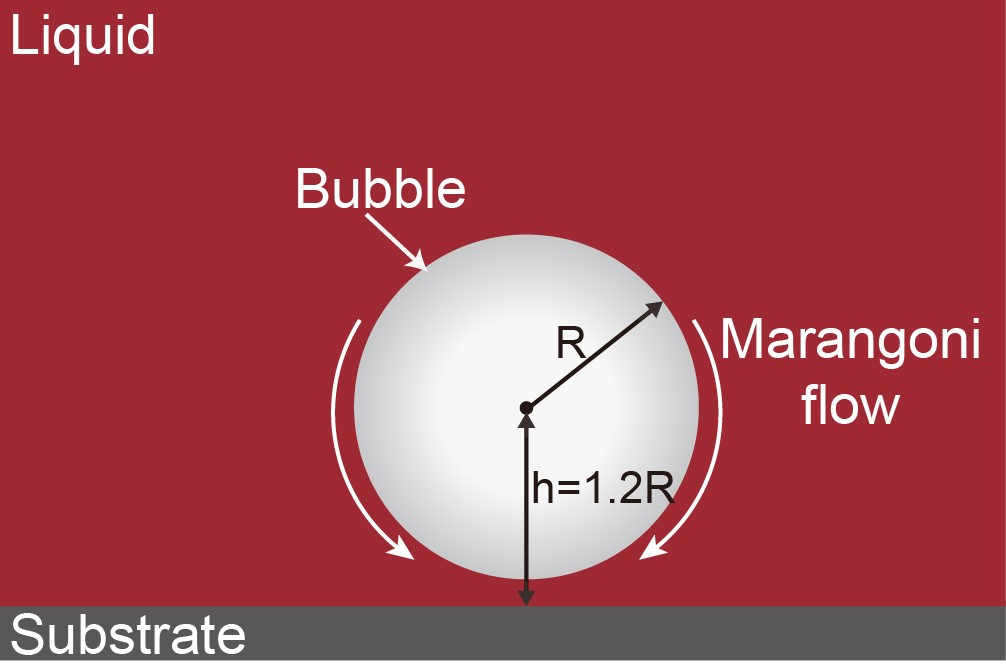}
  \caption{The numerical setup for a bubble with a constant radius $R$ fixed at $h = 1.2R$ above the bottom wall. There is a constant normal concentration gradient at the bubble surface due to the evaporation of ethanol in the surrounding ethanol water mixture. Marangoni flow is induced by the ethanol concentration difference along the bubble surface. Note that the simulation is conducted in a three dimensional domain although the sketch is two dimensional.}
\label{figc5s2}
\end{figure}

The simulation was conducted in a three dimensional domain $L_x\times L_y\times L_z = 20R\times 20R\times 20R$ with $201\times 201\times 201$ grids. A bubble with a radius $R$ is fixed at a distance $h = 1.2R$ above the middle of the bottom wall (Figure \ref{figc5s2}).

\begin{figure}[htbp]
\centering
  \includegraphics[width=0.85\textwidth]{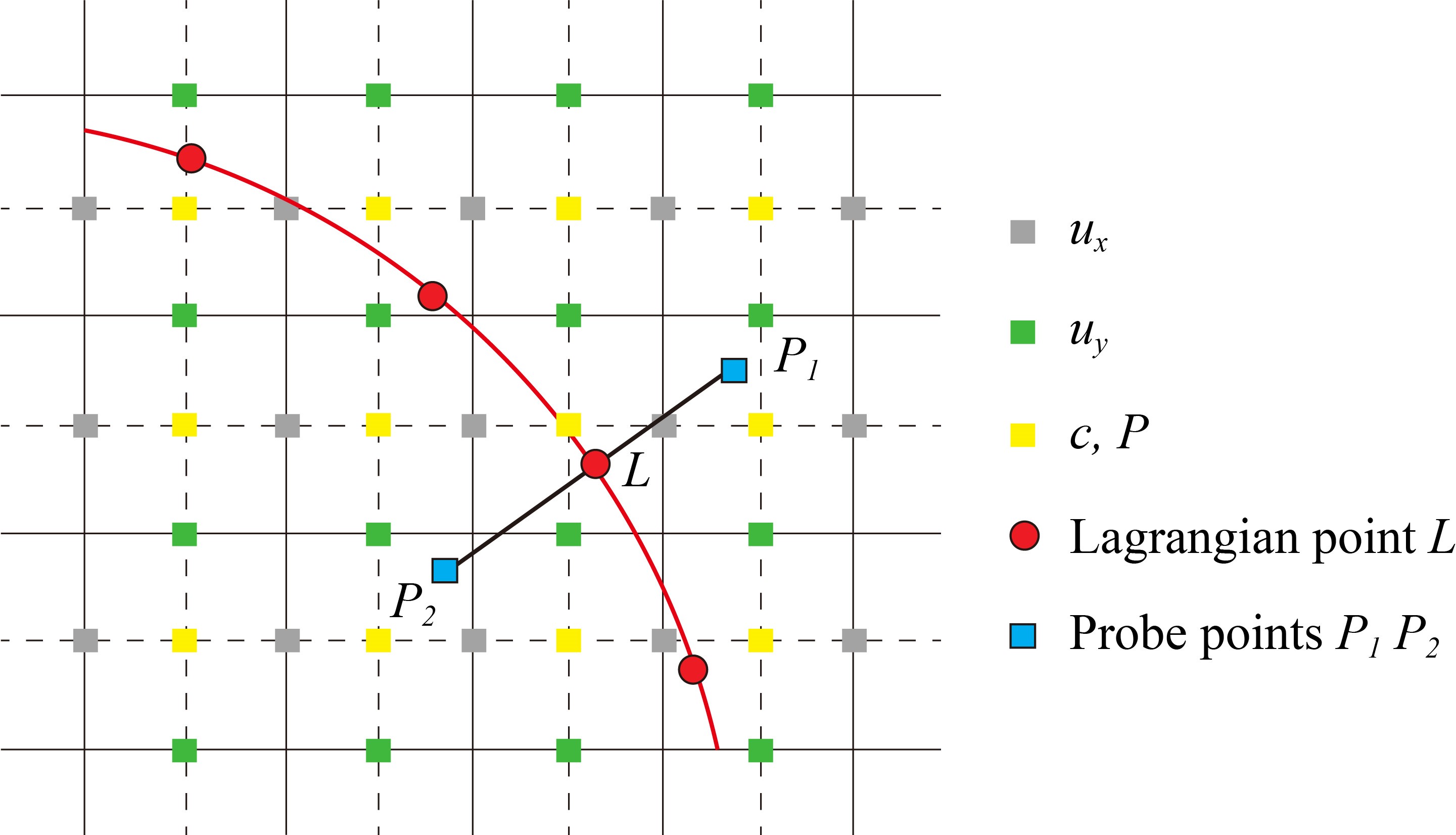}
  \caption{Schematic illustration of the immersed boundary method with staggered grids: Eulerian grids of velocity points and concentration points are denoted by different color, and the Lagrangian points at the boundary is denoted by red circle. The governing equations are solved on the Eulerian grids and the boundary conditions are satisfied at $L$ point based on the interpolated value at the probes $P_1$ and $P_2$ points.}
\label{figc5s3}
\end{figure}

In non-dimensional form the boundary conditions (Eqs. (\ref{equ1}) and (\ref{equ2})) read:
\begin{equation}
\partial_{\tilde{n}} \tilde{c}=1, \quad \partial_{\hat{\tau}} \tilde{c}=\partial_{\tilde{n}} \tilde{\boldsymbol{u}}
\label{eqs1}
\end{equation}
The governing equations of the flow field were the Navier-Stokes and diffusion-convection equations Eqs. (\ref{equ1}) and (\ref{equ2}). The equations were spatially discretized using the central second-order finite difference scheme. Along both horizontal and vertical directions, homogenous staggered grids were used. The equations were integrated by a fractional-step method with the non-linear terms computed explicitly by a low-storage third-order Runge-Kutta scheme and the viscous terms computed implicitly by a Crank-Nicolson scheme\cite{verzicco1996}.
To satisfy the concentration and velocity boundary conditions at the bubble surface, we applied the moving least square (MLS) based immersed boundary (IB) method, where the particle interface was represented by a triangulated Lagrangian mesh. The details of our MLS-based IB method are documented in Spandan \textit{et al}\cite{spandan2017}.
Schematic of Figure \ref{figc5s3} illustrates the immersed boundary method with staggered grids. For clarity, we use two-dimensional grids in Figure \ref{figc5s3} although the simulation is done in three dimensions. The Lagrangian points are denoted as red circle at the boundary (red curve), and the Eulerian points by the squares. A staggered mesh was applied to increase the accuracy. Therefore, the velocity points and concentration/pressure points are at different locations denoted by squares of different colors.

In the simulation, the concentration boundary condition (Eq. (\ref{eqs1})a) is satisfied by enforcing the concentration at the inner probe point (point $P_2$ in Figure \ref{figc5s3}) based on concentration interpolated at the probe located at a short distance (1 grid size h) from the surface of the particle (point $P_1$ in Figure \ref{figc5s3}):

\begin{equation}
\frac{c_{R}-c_{P_{2}}}{2 h}=1
\label{eqs2}
\end{equation}
The surface gradient of the concentration in the velocity boundary condition (Eq. (\ref{eqs1})b) was calculated by the nearby Lagrangian concentration. The velocity boundary condition at the surface of the particle (point $L$ in Figure \ref{figc5s3}) was then enforced based on the velocity interpolated at the probe located at a short distance (1 grid size) from the surface of the particle (point $P_1$ in Figure \ref{figc5s3}):
\begin{equation}
\frac{u_{R}-u_{L}}{h}=\nabla_{s} c_{L}
\label{eqs2}
\end{equation}
\clearpage
\section{Surface tension of ethanol-water mixtures}\label{e-w-mix}
\begin{figure*}[ht]
\begin{center}
    \includegraphics[width=0.6\textwidth]{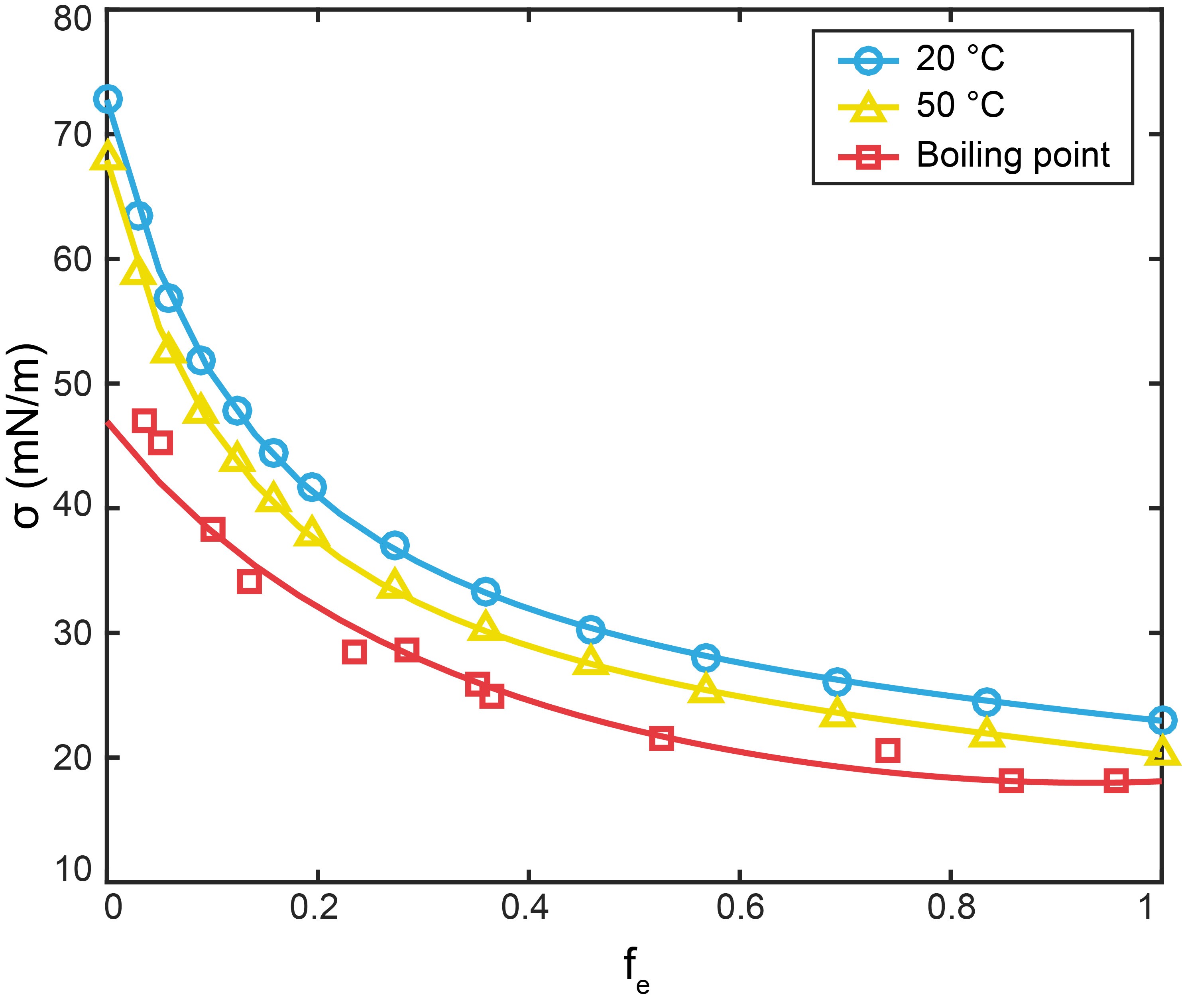}
    \caption{Surface tension of ethanol-water binary mixtures as a function of $f_e$ at 20 $^{\circ}$C, at 50 $^{\circ}$C, and at the respective boiling temperature of the mixture (which depends on $f_e$).}
    \label{s3}
    \end{center}
\end{figure*}

Figure \ref{s3} shows the surface tensions of the ethanol-water binary liquid with ethanol weight ratio $f_e$ from 0 to 1 at 20 $^{\circ}$C, at 50 $^{\circ}$C, and at boiling temperature (which depends on $f_e$). For each of the three curves, the surface tension $\sigma$ of the binary liquid decreases with the ethanol ratio $f_e$. It is also seen that $\sigma$ changes rapidly for small ethanol fractions and more slowly for larger ethanol fractions. The surface tensions for $f_e$ = 67.5$\%$ and $f_e$  = 80$\%$ are shown in Table \ref{table1}.
\begin{table}[ht]
 \caption{Surface tension of binary liquids with ethanol ratio $f_e$ = 67.5$\%$ and $f_e$ = 80$\%$}
 \centering
  \begin{tabular}{ccccc}
    \hline
    Temperature & $\sigma(f_e=67.5\%)$, mN/m & $\sigma(f_e=80\%)$, mN/m \\
    \hline
    20 $^\circ$C & 26.46  & 24.94 \\
    50 $^\circ$C & 23.8  & 22.31 \\
    Boiling point & 19.46  & 18.39\\
    \hline
  \end{tabular}
  \label{table1}
\end{table}
\clearpage
\section{Numerical calculated temperature field of the substrate}\label{temp}
\begin{figure}[h]
\centering
  \includegraphics[width=0.5\textwidth]{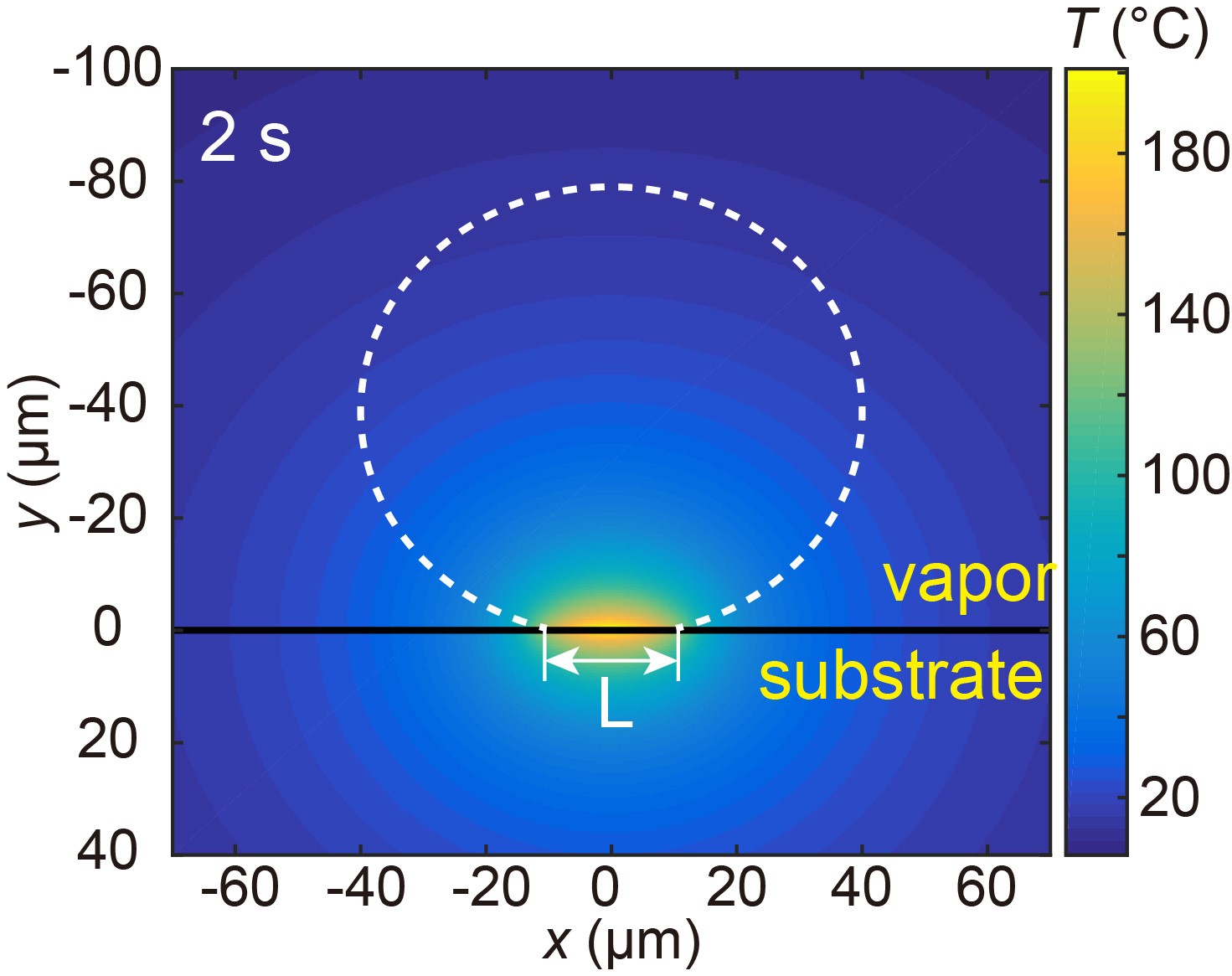}
  \caption{Numerical calculated temperature field of the substrate immersed in vapor upon laser irradiation for 2 s.}
\label{figc5s5}
\end{figure}

\end{appendix}

\bibliographystyle{prstywithtitle}

\bibliography{bibliopre}

\end{document}